%% file: main.tex
\documentclass{article}
\usepackage[utf8]{inputenc}
\usepackage{calligra}
\usepackage[T1]{fontenc}
\usepackage[pdftex]{hyperref}
\usepackage[bottom=2cm,top=2cm,left=2cm,right=2cm]{geometry}
\usepackage{amsmath}
\usepackage{amsfonts}
\usepackage{amssymb}
\usepackage{graphicx}
\usepackage{dsfont}
\graphicspath{ {./images/} }
\usepackage{chemfig}

\title{The vector MIT bag model under the light of new data.}
\author{Maíra Cesário Alvim Lobo, Débora Peres Menezes}
\date{January 2023}

\begin{document}

\maketitle

\begin{abstract}
    In the present work we use the vector MIT bag model to describe quark matter and obtain the macroscopic properties of neutron stars. 
    We also calculate the speed of sound of quark matter described by this model with specific parameter values and check if the results obey the conformal limit at high densities.
    We have seen that the vector MIT bag model produces stars with masses and radii consistent with recent data, suggesting the possible existence of massive quark (or strange) stars. 
\end{abstract}

\include{chapters/1-chapter}
\include{chapters/2-chapter}

\include{bibliography}
\bibliographystyle{unsrt}
\bibliography{references}




\end{document}

%% file: chapters/1-chapter.tex
\section{Introduction}

Deconfined quark matter is predicted by quantum chromodynamics (QCD) and is related to asymptotic freedom \cite{PhysRevLett.30.1343,PhysRevLett.30.1346}. The deconfinement of quark matter is expected to take place at high temperatures and low densities, high densities and low temperatures and other combinations of temperature and baryon densities connecting these two regions. Hence, deconfined matter may exist in the core of massive compact objects such as neutron stars.
At such high densities, present in the interior of these objects, the Bodmer-Witten conjecture states that deconfined strange quark matter (SQM) is not only possible, but it is probably the true ground state of the strong interaction  \cite{PhysRevD.4.1601,PhysRevD.30.272}. 

Since QCD is far from having an analytical solution, models are generally used to describe SQM and quark stars. The MIT bag model \cite{PhysRevD.9.3471,PhysRevD.10.2599,PhysRevD.12.2060} introduced in the 70's is well known and has been modified more recently to include a repulsive interaction \cite{modifiedmitbagmodel} to account for the description of massive neutron stars detected in the last decade or so. It is called vector MIT bag model.{  The inclusion of vector channels in the MIT bag model is nothing new, as can be seen in \cite{Gomes_2019, Gomes_2019_2,Klähn_2015,Cierniak_2019,Wei_2019}, but the vector MIT bag model first introduced in \cite{modifiedmitbagmodel} and used in the present work has the great advantage to be able to produce stars with masses and radii consistent with the new data and also produce a QCD phase diagram that obeys experimental constraints and approximate ranges for the  critical chemical potential that is inferred from other models and calculations. By critical chemical potential we mean the critical chemical potential as the chemical potential at which the phase transition occurs at zero temperature ($T=0$). To be consistent with experimental data and the range inferred from other models and calculations with the same model parameters, another change in the model was introduced in \cite{parteii} so that the bag pressure depends on the temperature. However, for the temperature range of $T=0$ MeV to $T=40$ MeV, used in this paper, this change has no significant effect (at most 1,001\% on the value of the bag constant) so this dependence is not considered here. It is important to point out that the parameters that produce star families comprising the new data are not the same parameters that produce QCD phase diagrams that obey the constraints considered, as discussed in \cite{parteii}.
 
After the paper \cite{modifiedmitbagmodel} was published, new neutron stars (NS) data were sent from the NICER (Neutron star Interior Composition ExploreR) telescope, a mission supported by NASA, as the mass and radius of the massive pulsar PSR J0740+6620 \cite{Riley_2021}   and the radius of the canonical star\cite{capano} .
Another interesting pulsar is a black widow known as PSR J0952-0607, whose
data were obtained using spectrophotometry and imaging from the Keck Telescope \cite{psrj0952} and it is possibly the most massive and fastest object ever observed. 
The purpose of the present paper is to analyse whether the vector MIT bag model describes these recently observed compact objects.

In addition, we also calculate the speed of sound, which is an important quantity because according to \cite{soundspeed}, ``for the heaviest reliably observed neutron stars with masses $M\approx 2M_{\odot}$ the presence of deconfined quark matter is associated with the behaviour of the speed of sound $v_s$ in strongly interacting matter. If the conformal bound $v_s^2\le 1/3$ is not strongly violated, massive neutron stars are predicted to have sizable quark matter cores.'' The conformal limit is related to the condition that the speed of light is greater than the speed of sound in a medium.

In the next section, we first outline the main expressions necessary to understand the formalism, and then explain which quark coupling constants are used. We then obtain and discuss the results.

%% file: chapters/2-chapter.tex
\section{Vector MIT bag model}
The original MIT bag model describes a relativistic gas of  
quarks inside a volume delimited by a bag. Its Lagrangian density describes a relativistic gas of free fermions plus a bag constant ($\mathbf{\mathcal{B}}$) that represents the pressure needed to balance the degeneracy pressure of the quarks. A Heaviside function is introduced to ensure the confinement. The vector MIT bag model is a modification of this model that introduces a mesonic vector field that mimics a repulsive interaction in the system. 
We use the mean field approximation (MFA)
\begin{equation*}
    V^{\mu}\;\rightarrow\;\langle V \rangle\; \rightarrow\; \delta_{0,\mu}V^0.
\end{equation*}massive
and within this approximation, the Lagrangian density of the model reads:
\begin{equation}\
    \mathbf{\mathcal{L}}=\sum_q\Big\{\overline{\psi}_q[\gamma^{\mu}(i\partial_{\mu}-{\color{blue}g_{qqV}\;\delta_{\mu,0}V_0})-m_q]\psi_{q}+{\color{blue}\frac{1}{2}m_{V}^2V_{0}^2+\frac{b_4(g_{uuV}V_0)^4}{4}}-\mathbf{\mathcal{B}}\Big\}\theta(\overline{\psi}_q\psi_q)
\end{equation}
where the index $q$ relates to the quark flavors (up, down and strange), $m_q$ are the quark masses (we use $m_u=m_d=4$ MeV and $m_s=95$ MeV), $\psi_q$ are the quark fields, $g_{qqV}$ are the quark coupling constants, $m_{V}$ is the meson mass and $b_4$ is the self-interacting parameter. The parameter values are discussed ahead. The blue terms indicate the modifications introduced in \cite{modifiedmitbagmodel}  while the black parts show the Lagrangian density of the original MIT bag model. 

The equation of motion obtained from the Euler Lagrange equation
is
\normalsize{\begin{equation}
\label{termocomv0}
g_{qqV}\;V_0\;\rho_q=m^2_V\;V_0^2+b_4\;(g_{uuV}\;V_0)^4.
\end{equation}}
 
We derive the equation of state (EOS) from the model Lagrangian using thermodynamics and statistics theory. To obtain the pressure and the energy density we use the energy-momentum tensor($\varepsilon=\langle T_{00}\rangle$ and $P=\frac{1}{3}\langle P_{ii}\rangle$) $T_{\mu\nu}$ and the baryonic density is the sum of the quark densities. The EOS of the vector MIT bag model is given by:

\begin{equation}
\label{eosp}
P=\sum_q\Big\{\frac{1}{\pi^2}\int dp_q \frac{p_q^4}{\sqrt{p_q^2+m_q^2}}\;(f_{q+}+f_{q-}){\color{blue}+\frac{m_V^2\;V_0^2}{2}+\frac{b_4(g_{uuV\;V_0})^4}{4}}\Big\}-\mathbf{\mathcal{B}}
\end{equation} 
\begin{equation}
\label{eostempvett}
    \varepsilon=\sum_q\Big\{\frac{3}{\pi^2}\int dp_q\; p_q^2\;\sqrt{p_q^2+m_q^2}\;(f_{q+}+f_{q-}){\color{blue}+\frac{m^2_V\;V_0^2}{2}+\frac{3b_4\;(g_{uuV}\;V_0)^4}{4}}\Big\}+\mathbf{\mathcal{B}}
\end{equation}
\begin{equation}
    \rho_{B}=\frac{1}{3}\sum_q\frac{3}{\pi^2} \int dp_q\; p_q^2\; (f_{q+}-f_{q-})
\end{equation}
where $p_q$ is the quark momentum and  $f_{\pm}$ are respectively the particle and antiparticle distribution functions:
\begin{equation}
    f_{q\pm}=\frac{1}{1+e^{(\epsilon_q(\mathbf{p})\mp\mu_q)/{T}}}
\end{equation}
for finite temperature and 
\begin{equation}\label{dist-theta}
    f_{q+}=\theta(p_q-p_{qF})\;\;\;\;\text{and}\;\;\;\; f_{q-}=0.
\end{equation}
for $T=0$, where $p_{qF}$ is the Fermi momentum of quark $q$.

The free parameters of the model are:

\begin{equation*}
    {\mathbf{\mathcal{B}}},\;\;\;G_V=\Big(\frac{g_{uuV}}{m_V}\Big)^2 ,\;\;\; X_V=\frac{g_{ssV}}{g_{uuV}}\;\;\;\text{and}\;\;\;  b_4 .
\end{equation*}
In this paper, we use exclusively $G_V=0.3$ fm$^2$, $b_4=-0.4$ and two possibilities for $X_V:\;X_V=1.0$ and $X_V=0.4$.We restrict ourselves to these values of $G_V$ and $b_4$ because they generate stiffer EOS and consequently bigger star masses, as can be seen in \cite{modifiedmitbagmodel}. For a better understanding of the calculation and values chosen for the quark coupling constants, the interested reader can look at appendix A of \cite{modifiedmitbagmodel}, where the SU(3) and SU(6) symmetry groups are used. The interval of values for the bag constant are presented in the next section.

\subsection{Stellar matter}

The Bodmer-Witten conjecture is based on the idea that, at high density, the presence of the strange quark turns the deconfined quark matter into the true ground state of the strong interaction by lowering its binding energy. This conjecture can be summarized by ensuring that the binding energy of iron is lower than the binding energy of $u$ and $d$ quark matter and higher than the binding energy of the strange quark matter ($u$, $d$ and $s$ included).
These constraints when imposed upon the model's parameters, give us the stability window of the model, i.e., the range of the parameters values that describe a stable strange quark matter. From \cite{modifiedmitbagmodel} we have that, for the parameter values that we use here, the stability window is as shown in table \ref{tab:stabilitywindow}. It is worth mentioning that the parameter $b_4$ doesn't have any influence on the obtained values. {So for describing a star with stable matter, we must use  the parameters within the stability window.}
\begin{table}[]
    \centering
    \begin{tabular}{|c|c|c|c|}\hline
{$G_V$ (fm$^2$)} & $X_V$ & Minimum value of $\mathbf{\mathcal{B}^{1/4}}$ & Maximum value of $\mathbf{\mathcal{B}^{1/4}}$\\ \hline
0.3                                 & 1.0   & 139                                                              & 146                                                              \\\hline
0.3                                 & 0.4   & 139                                                              & 150                                                            \\ \hline
\end{tabular}
    \caption{Stability window of Vector MIT bag model.}
    \label{tab:stabilitywindow}
\end{table}

Stellar matter comprises leptons (here we use $e^-$ and $\mu$)
and quarks because of the necessary charge neutrality and $\beta$ equilibrium, conditions given respectively by:

\normalsize{\begin{equation}\label{neutralidade elétrica}
    \rho_{e}+\rho_{\mu}=\frac{1}{3}(2\rho_{u}-\rho_{d}-\rho_{s}),
\end{equation}}
and 
{\begin{equation}\label{equilibriobeta}
    \mu_{d}=\mu_{s}=\mu_{u}+\mu_{e},\;\;\; \mu_{e}=\mu_{\mu}.
\end{equation}}

Hence, we also have to consider the equation of state of the leptons, which is similar to the EOS of the quarks, but neither the bag constant nor the vector interaction are present and divided by three because of the absence of color degeneracy.

To generate neutron star families using a model EOS, the Tolman-Oppenheimer-Volkoff (TOV) equations \cite{PhysRev.55.364,PhysRev.55.374} are necessary. They relate the microphysics (the matter that constitutes the star) with the macrophysics (radius and mass, for instance) of the star. They are built in the context of general relativity using Einstein's equations considering the star spherically symmetric, static, and as a perfect fluid. In the natural units system, the TOV equations are
given by
\begin{equation}
    \frac{dP(r)}{dr}=-\frac{[P(r)+\varepsilon(r)][M(r)+4\pi r^3P(r)]}{r[r-2GM(r)]}
\end{equation}
and 
\begin{equation}
    \frac{dM}{dr}=4\pi r^2 \varepsilon(r) 
\end{equation}
where $\epsilon$, $P$ and the star mass $M$ are evaluated  in the layer of radius R.

\begin{figure}
\centering
\begin{minipage}{.5\textwidth}
  \centering
  \includegraphics[width=1.0\linewidth]{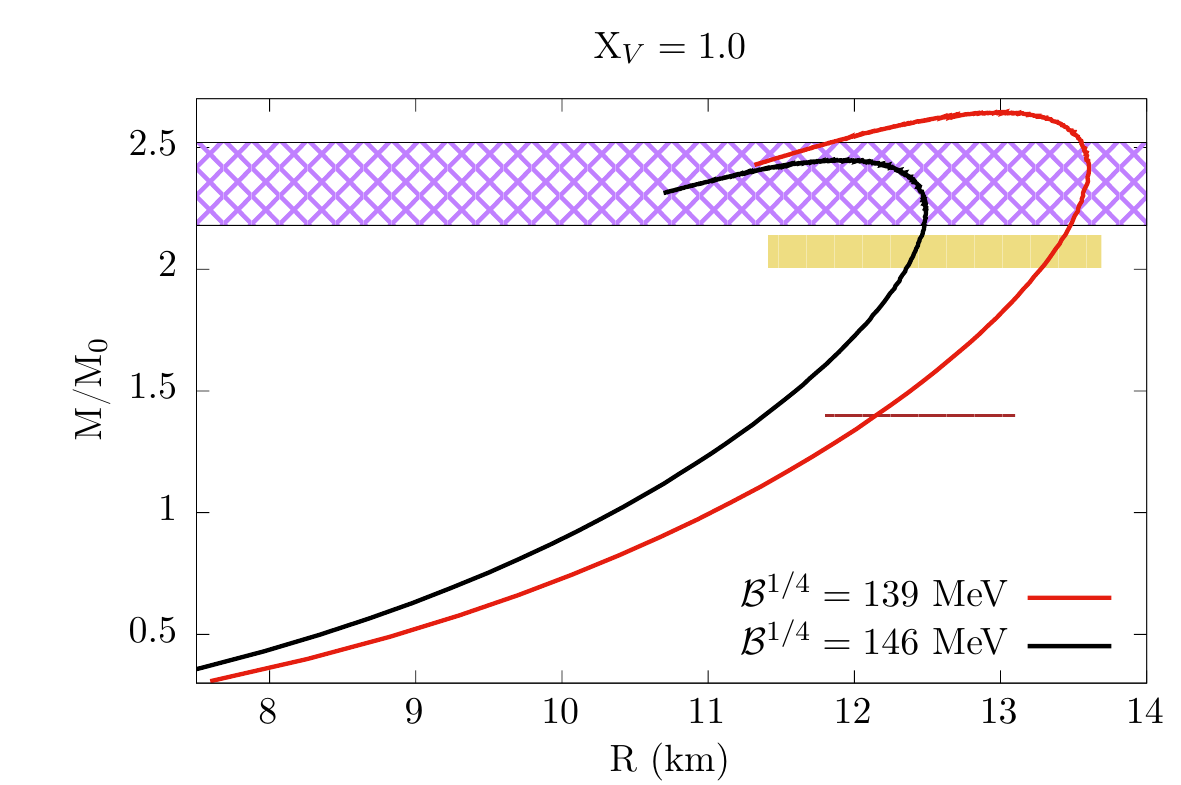}
\end{minipage}%
\begin{minipage}{.5\textwidth}
  \centering
  \includegraphics[width=1.0\linewidth]{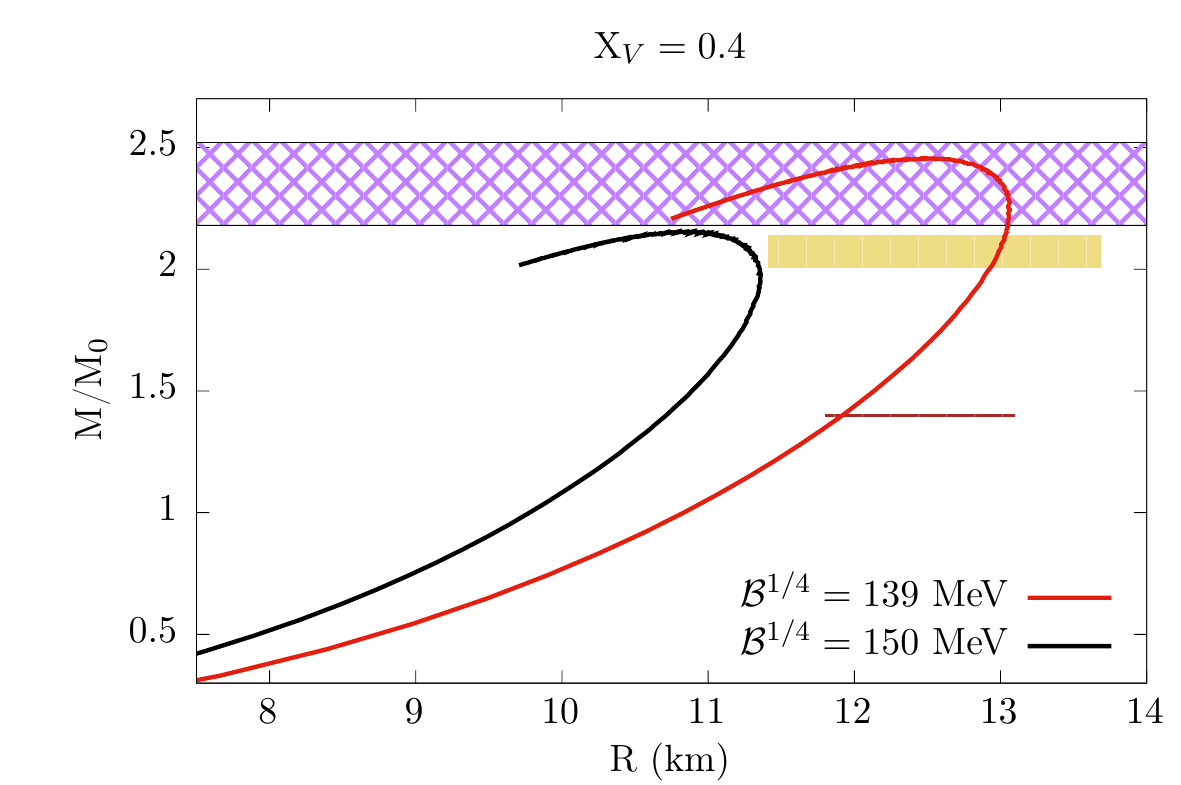}
\end{minipage}
\caption{ Mass radius diagrams generated byuse of the EOS of vector MIT bag model with the bag constants within the values of the stability window extremes \ref{tab:stabilitywindow}, $G_V=0.3$ fm$^2$, $b_4=-0.4$ and $X_V=1.0$ (left) and $X_V=0.4$ (right).
\\
1- Rectangle with the lilac checkered pattern: PSR J0952-0607 \cite{psrj0952} (2022).\\ 
2- Yellow rectangle: PSR J0740+6620 \cite{Riley_2021}  (2021)\\
3- Brown line: canonical star radius \cite{Riley_2021}  (2021).}
\label{TOV}
\end{figure}

With the stability window in mind, we can study the masses and radii of stable strange stars for every parameter set { shown in Table 1}. In Figure \ref{TOV} the TOV \cite{PhysRev.55.364,PhysRev.55.374} solutions for the minimum and maximum allowed bag pressure values that produce stable strange quark matter are depicted. In these mass-radius diagrams, there are data from \cite{Riley_2021}, \cite{capano} and \cite{psrj0952}. The data from NICER \cite{Riley_2021} of the pulsar PSR J0740+6620 with its 
error (M=$2.072^{+0.067}_{-0.066}$ M$_{\odot}$ and R=$12.39^{+1.3}_{-0.98}$ km) is represented as the yellow rectangle. The data for the interval of the canonical star radius (R=$12.45\pm 0.65$), from \cite{capano}, is expressed as the brown line where the canonical star is the star with mass M=1.4 M$_{\odot}$. The new data from the pulsar PSR J0952-0607 \cite{psrj0952} (M=$2.35\pm 0.17$ M$_{\odot}$) is also plotted as the rectangle with the lilac checkered pattern. {As mentioned in the Introduction, PSR J0952-0607 is a black widow and its mass has a huge error bar hence, these values have to be taken with care.}  

 A desirable solution of the TOV equations describes both the data of the canonical star and the data of the massive objects mentioned above.  In Table \ref{tab:massasmaximas} the parameters used are displayed in the first and second columns.
In the third column, the maximum mass of the star family is shown and the sixth and seventh column show if it reaches the mass and radius of PSR J0740+6620 and PSR J0952-0607 respectively. In the fourth column, the radius of the canonical star is shown and the fifth column shows if it fits the data for the radius of the canonical star. Analysing Figure \ref{TOV} and Table \ref{tab:massasmaximas} we see that, with an appropriate choice of values, the vector MIT bag model describes the pulsars 
analysed in this paper. It is noticeable from figure \ref{TOV} that for $X_V=1.0$ more values for the bag constant, within the stability window, generate families of stars that describe the new data. 

 \begin{table}[]
     \centering
     \begin{tabular}{|c|c|c|c|c|c|c|c|}\hline
        $\mathbf{\mathcal{B}}^{1/4}$ (MeV) & $X_V$ & M$_{max}$ & R$_{1.4}$ km & Within R$_{1.4}$  & {Within PSR J0740+6620  } & Within PSR J0952-0607 \\
                                           &       & (M$_{\odot}$)   &              & constrains?     &     constrains?       &   constrains?                 \\ \hline
        139             & 1.0   & 2.65      &   12.15      & YES         & YES                  & YES                \\ 
        146             & 1.0   & 2.45      &   11.40      & NO         & YES                  & YES                \\ \hline 
        139             & 0.4   & 2.45      &   11.92      & YES         & YES                  & YES                \\ 
        150             & 0.4   & 2.15      &   10.71      & NO           & NO                  & NO                \\ \hline 
        
        \end{tabular}
     \caption{Data on star families of vector MIT bag model for $G_V=0.3$ fm$^2$ and $b_4=-0.4$. 
     }
     \label{tab:massasmaximas}
 \end{table}


 \begin{figure}
     \centering
     \includegraphics[scale=0.2]{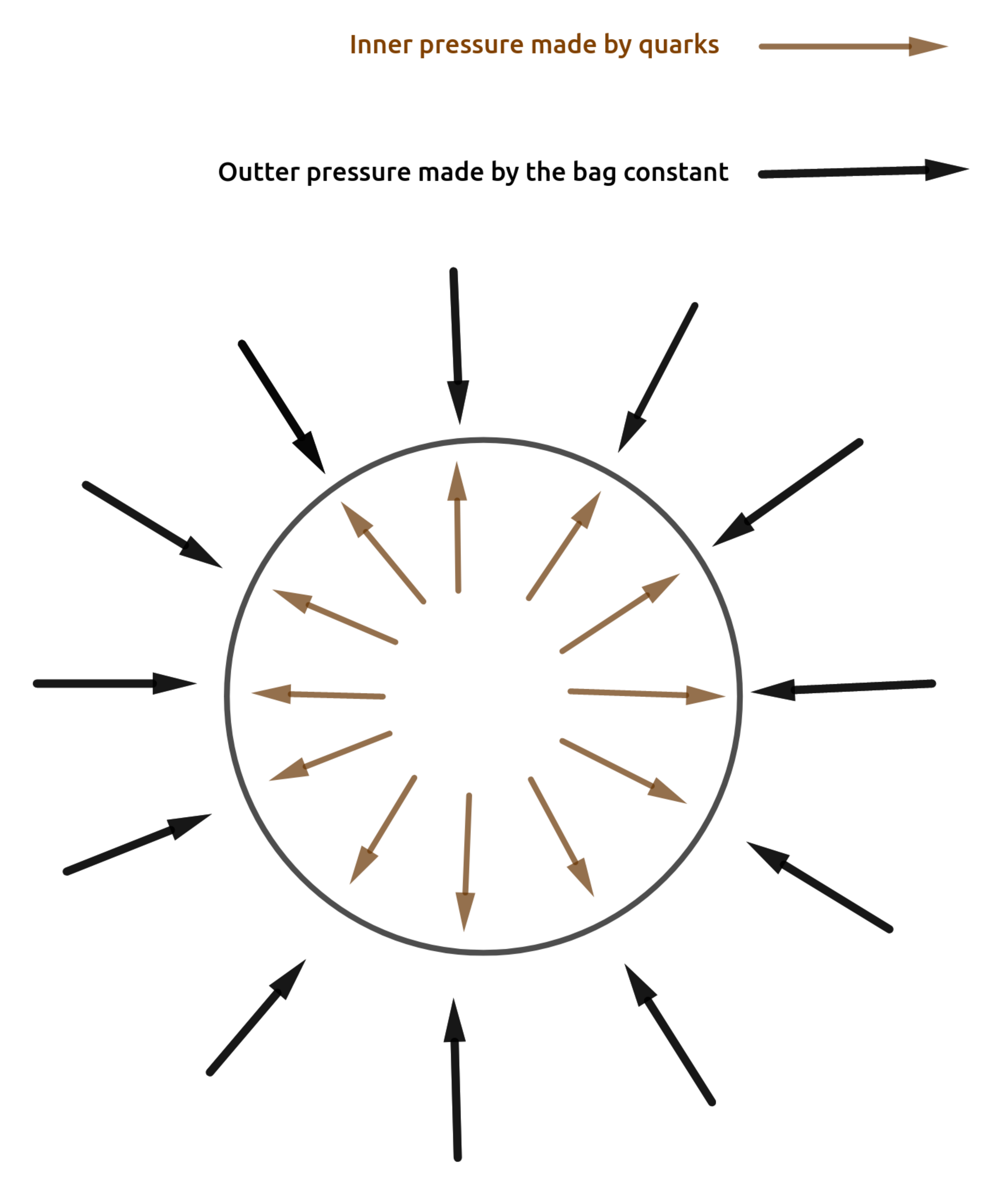}
     \caption{Illustration of the balance of the pressure on the bag.}
     \label{fig:pressure}
 \end{figure}

\begin{figure}
    \centering
    \includegraphics{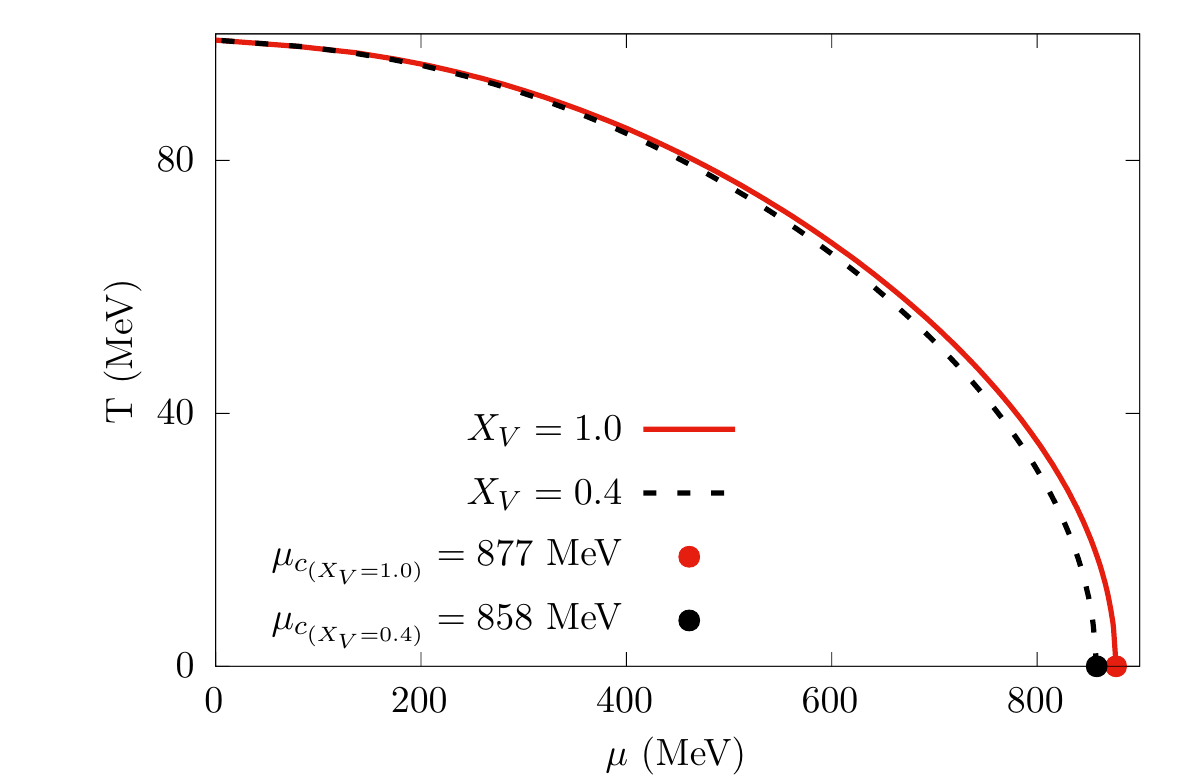}
    \caption{Phase diagrams for $G_V=0.3$ fm$^{2}$, $b_4=-0.4$, $\mathbf{\mathcal{B}}=139$ MeV} and both values of $X_V$.
    \label{fig:phasediagram}
\end{figure}

 {Now we consider the phase transition diagram , from confined to deconfined matter, for quark matter generated by using the vector MIT bag model.  The pressure on the bag is illustrated in figure \ref{fig:pressure}.} A naive prescription to estimate the phase transition from confined to deconfined matter is to calculate the chemical potential at different temperatures with the condition of zero pressure, {which corresponds to the Gibbs' conditions of phase coexistence.
The prescription is based on the idea that the pressure of the quarks plus the pressure of the vector mesons inside the bag equals the bag constant in such a way that the bag no longer supports the inner pressure, giving rise to a phase coexistence and ultimately, for slightly larger bag values, deconfinement.}
We plot a quark matter phase diagram of the vector MIT bag model in Fig.\ref{fig:phasediagram} for both values of $X_V$ and $\mathbf{\mathcal{B}^{1/4}}=139$ MeV. This bag constant is chosen because it generates star families with greater masses 
as seen in \cite{modifiedmitbagmodel}. In the diagram we have the information of the critical chemical potential $\mu_c$, that is the chemical potential where the phase transition occurs when $T=0$ MeV.

{A different scenario is also possible, in which the core of the neutron star undergoes a phase transition to quark matter, while the outer layers are still formed by hadronic matter.
These conditions are the ones found in hybrid stars, proposed for the first time in 1969 \cite{quarkstars}.
A recent study \cite{soundspeed} suggests that the hybrid star is indeed the most probable scenario for massive neutron stars.} As pointed out in \cite{soundspeed}, the speed of sound in quark matter can be related to the mass and radius of the quark core of deconfined quark matter in hybrid stars. The authors found that if the conformal limit
 \begin{equation*}
     v^2_s=\frac{\partial p}{\partial \varepsilon}<\frac{1}{3}
 \end{equation*}
 is not strongly violated then massive neutron stars are likely to have sizable quark-matter cores,
 greater than $0.25$ M$_{\odot}$ and reaching $0.8$ M$_{\odot}$ \cite{soundspeed}, corresponding respectively to 9.4\% and 30.2\% of the total star mass (2.65 M$_{\odot}$).
 
\begin{figure}
     \centering
     \includegraphics{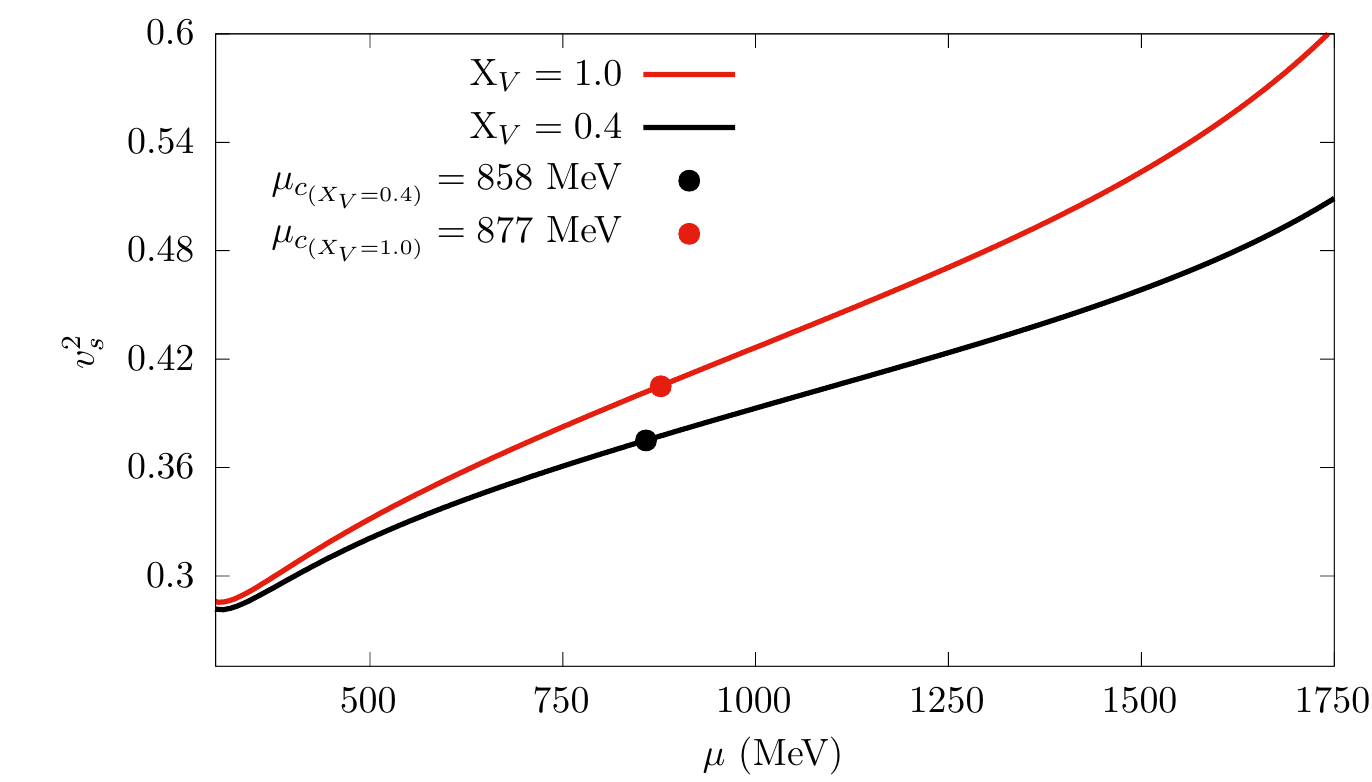}
     \caption{The square of sound speed for $G_V=0.3$ fm$^{2}$, $b_4=-0.4$ and both values of $X_V$.}
     \label{fig:soundspeed}
 \end{figure}

We then analyse the speed of sound calculated with the models presented here to check how far/close it is with respect to the conformal limit.
The speed of sound as a function of the chemical potential is plotted in Fig.~\ref{fig:soundspeed}. Note that it does not depend on the bag constant. The chemical critical potentials were calculated and shown in the figure by the red (for $X_V=1.0$) and black (for $X_V=0.4$) dots.
We see that the squared speed of sound at $\mu_c$ is always higher than
 $1/3$ and presents an increasing behavior with the increase of the chemical potential. Had we chosen this specific model to describe quark matter inside hybrid stars, the probable result would be a small core.
 However, there is no reason to choose values that satisfy the stability window in the description of hybrid stars. Actually, the opposite choice is desirable to avoid conversion from a hybrid to a strange star \cite{kauan,lopes2022hypermassive},  and the sound velocity would have to be reobtained with other parameter values. In \cite{parteii} one can see that for certain parameter sets the conformal limit is respected.
 
 For the sake of completeness, the  value of the  squared speed of sound  at the critical chemical potential, the maximum star mass, and information about the radius of the canonical star (both sets of parameters agree with data for the radius of the canonical star) are given in table \ref{tab:data}. The masses on the Table encompass both pulsars  PSR J0952-0607 and PSR J0740+6620.

 \begin{table}[]
     \centering
     \begin{tabular}{|c|c|c|c|c|}\hline 
         $X_V$ & $\mu_{c}$ (MeV) & $v_{s}^2$ at $\mu_c$ & M$_{max}$ (M$_{\odot}$) & $R_{1.4\odot}$ (km) ?\\\hline 
          1.0  & 877             & 0.41                 & 2.65                    &  YES                 \\ \hline 
          0.4  & 858             & 0.38                 & 2.45                    &  YES                 \\ \hline 
     \end{tabular}
     \caption{Data for the vector MIT bag model with $G_V=0.3$ fm$^2$, $b_4=-0.4$ and $\mathbf{\mathcal{B}}=139$ MeV.
     The fifth column says both sets of parameter generate stars that satisfy the data from canonical star radius.}
     \label{tab:data}
 \end{table}

 \section{Final remarks}

  In this paper, we revisited the vector MIT bag model, which is an extension of the well-known MIT bag model. The modifications concern to the introduction of a mesonic vector field that acts as a source of interaction between the quarks. The free parameters of the model are associated with the bag constant value, the vector field intensity, the quark coupling constants and the self-interacting field.
We have partially fixed the quark coupling constants by satisfying symmetry group theory requirements, as done in \cite{modifiedmitbagmodel}.

To study quark stars, stellar matter is defined as a neutral matter composed of quarks and leptons ($e^{-}$ and $\mu$ in this work). This kind of matter has to be in $\beta$ equilibrium.
Based on the Bodmer-Witten conjecture, we checked when stellar matter is stable by obtaining the stability window, {that is an interval for the model parameters in which strange matter is stable}.
The TOV equations were introduced to relate the microphysics of the quark matter (the EOS) to the macrophysics of the star, as its mass and radius. 
{ By comparing the results of the TOV equations when using the EOS of the vector MIT bag model with recent data from compact stars, PSR J0952-0607 (2022) and PSR J0740+6620 (2021),  we conclude that this model is efficient in describing the new findings. }

After a discussion about quark phase transition in MIT-based models, the possibility of hybrid stars was mentioned. As pointed out in \cite{soundspeed}, the speed of sound in quark matter can give an estimation of the quark core size. This quantity was then calculated and  analyzed for the vector MIT bag model. The speed of sound in the stars generated by the set of parameters that agrees with the recent data does not obey the conformal limit \cite{soundspeed}.

 We conclude that using the right sets of parameters the vector MIT bag model introduced in \cite{modifiedmitbagmodel} can describe the recent data of masses and radii of the pulsars PSR J0952-0607 (2022) \cite{psrj0952} and  PSR J0740+6620 (2021) \cite{Riley_2021}.

\hspace{5cm}

{\bf Acknowledgments} 

This work is a part of the project INCT-FNA Proc. No. 464898/2014-5. D.P.M. was partially supported by Conselho Nacional de Desenvolvimento Científico e Tecnológico (CNPq/Brazil) under grant 303490-2021-7  and  M.C.A.L. acknowledges a doctorate scholarship from Coordenação de Aperfeiçoamento de Pessoal do Ensino Superior (Capes/Brazil).  M.C.A.L. thanks fruitful discussions with Carline Biesdorf.